\documentclass[aps,prl,twocolumn,showpacs,superscriptaddress,groupedaddress]{revtex4}  
\usepackage{graphicx}  
\usepackage{dcolumn}   
\usepackage{bm}        
\usepackage{amssymb}   
\usepackage{amsmath}
\usepackage{pifont}    
\usepackage{color}
\usepackage{umoline}

\makeatletter 
\@ifundefined{textcolor}{}
{%
 \definecolor{BLACK}{gray}{0}
 \definecolor{WHITE}{gray}{1}
 \definecolor{RED}{rgb}{1,0,0}
 \definecolor{GREEN}{rgb}{0,1,0}
 \definecolor{BLUE}{rgb}{0,0,1}
 \definecolor{CYAN}{cmyk}{1,0,0,0}
 \definecolor{MAGENTA}{cmyk}{0,1,0,0}
 \definecolor{YELLOW}{cmyk}{0,0,1,0}
 }

\begin{document}
\widetext
\title{Detecting and polarizing nuclear spins with double resonance
on a single electron spin}

%
\affiliation{Department of Physics, Technion, Israel Institute of
Technology, Haifa, 32000, Israel} \affiliation{Institut f\"{u}r
Quantenoptik, Universitat Ulm, 89073 Ulm, Germany}
\affiliation{Institut f\"{u}r Theoretische Physik, Albert-Einstein
Allee 11, Universitat Ulm, 89069 Ulm, Germany} \affiliation{Racah
Institute of Physics, The Hebrew University of Jerusalem,
Jerusalem, 91904, Israel} \affiliation{Graduate School of Library,
Information and Media Studies, University of Tsukuba, 1-2 Kasuga,
Tsukuba, Ibaraki 305-8550, Japan} \affiliation{National Institute
for Materials Science, Tsukuba, Ibaraki 305-0044, Japan}

\author{P. London}\affiliation{Department of Physics, Technion, Israel Institute of Technology, Haifa, 32000, Israel}
\author{J. Scheuer}\affiliation{Institut f\"{u}r Quantenoptik, Universitat Ulm, 89073 Ulm, Germany}
\author{J.-M. Cai}\affiliation{Institut f\"{u}r Theoretische Physik, Albert-Einstein Allee 11, Universitat Ulm, 89069 Ulm, Germany}
\author{I. Schwarz}\affiliation{Institut f\"{u}r Theoretische Physik, Albert-Einstein Allee 11, Universitat Ulm, 89069 Ulm, Germany}
\author{A. Retzker}\affiliation{Racah Institute of Physics, The Hebrew University of Jerusalem, Jerusalem, 91904, Israel}
\author{M.B. Plenio}\affiliation{Institut f\"{u}r Theoretische Physik, Albert-Einstein Allee 11, Universitat Ulm, 89069 Ulm, Germany}
\author{M. Katagiri}\affiliation{Graduate School of Library, Information and Media Studies, University of Tsukuba, 1-2 Kasuga, Tsukuba, Ibaraki 305-8550, Japan} \affiliation{National Institute for Materials Science, Tsukuba, Ibaraki 305-0044, Japan}
\author{T. Teraji}\affiliation{National Institute for Materials Science, Tsukuba, Ibaraki 305-0044, Japan}
\author{S. Koizumi}\affiliation{National Institute for Materials Science, Tsukuba, Ibaraki 305-0044, Japan}
\author{J. Isoya}\affiliation{Graduate School of Library, Information and Media Studies, University of Tsukuba, 1-2 Kasuga, Tsukuba, Ibaraki 305-8550, Japan}
\author{R. Fischer}\affiliation{Department of Physics, Technion, Israel Institute of Technology, Haifa, 32000, Israel}
\author{L. P. McGuinness}\affiliation{Institut f\"{u}r Quantenoptik, Universitat Ulm, 89073 Ulm, Germany}
\author{B. Naydenov} \affiliation{Institut f\"{u}r Quantenoptik, Universitat Ulm, 89073 Ulm, Germany}
\author{F. Jelezko} \affiliation{Institut f\"{u}r Quantenoptik, Universitat Ulm, 89073 Ulm, Germany}

\begin{abstract}
We report the detection and polarization of nuclear spins in
diamond at room temperature by using a single nitrogen-vacancy
(NV) center. We use Hartmann-Hahn double resonance to coherently
enhance the signal from a single nuclear spin while decoupling
from the noisy spin-bath, which otherwise limits the detection
sensitivity. As a proof-of-principle we: (I) observe coherent
oscillations between the NV center and a weakly coupled nuclear
spin, (II) demonstrate nuclear bath cooling which prolongs the
coherence time of the NV sensor by more than a factor of five. Our
results provide a route to nanometer scale magnetic resonance
imaging, and novel quantum information processing protocols.
\end{abstract}

\maketitle
Measurements of nuclear spin moments are essential to numerous
fields including medicine \cite{MansfieldNobel2004}, chemistry
\cite{Slichter}, metrology \cite{SimpsonGyro1963}, and quantum
information processing (QIP) \cite{NeumannScience2008}. Within
these, detection and manipulation of single or few nuclear spins
may revolutionize microscopy of biological systems with the
possibility to reveal the structure of single molecules. Moreover,
the potential of single nuclear spins as long-lived quantum memory
units is of intense current interest \cite{MaurerScience2012}.

However, measurements on single or small ensembles ($<$10$^{3}$)
of nuclear spins are extremely challenging due to the small
nuclear magnetic moment, leading to typically low polarizations,
especially at room temperature. Essentially, one must employ a
probe close enough to establish the required sensitivity, since
the coupling of the probe and the target spin decreases with the
distance between them. So far these have only been achieved with
magnetic resonance force microscopy \cite{Sidles1995}, quantum
dots \cite{Greilich2007}, and recently with the nitrogen-vacancy
(NV) center in diamond \cite{JelezkoPRL2004,ChildressScience2006}.
The NV center is an attractive system for this task: its optical
polarization and spin-dependent photoluminescence along with long
ground-state coherence time, make it a perfect probe for sensing
nuclear spins coupled to it via dipole-dipole interaction
\cite{DuttScience2007}.

As the large background noise originating from the spin-bath makes
dynamical decoupling techniques a necessity \cite{Taylor2008}, the
optimal way to uncover the target signal is not yet fully clear.
Recently, three studies have demonstrated the use of pulsed
dynamical decoupling to isolate the signal of a single nuclear
spin from the nuclear bath
\cite{Kolkowitz2012,Taminiau2012,Zhao2012}. Other complementary
techniques, applied to small ensembles, have observed statistical
fluctuations of nuclear spin states
\cite{Mamin2013,Staudacher2013}. These signals can be greatly
enhanced by hyperpolarization of the nuclei, if such is at
disposal.

Here we experimentally show that one can use \emph{continuous
dynamical decoupling} (CDD) \cite{Facchi2004,Fanchini2007,Cai1} to
overcome both challenges, namely to separate a single nuclear spin
signal from the bath noise, and to actively enhance the nuclear
polarization of the surrounding bath. In CDD, one applies a
continuous, resonant field to isolate the driven spin sensor from
its environment. The sensor spin is then insensitive to the
surrounding spins, however specific frequency components can be
selected through a phenomenon known in nuclear magnetic resonance
as Hartmann-Hahn double resonance (HHDR) \cite{Hartmann1963,Cai1}.
We use this technique to experimentally implement an imaging
scheme recently proposed in reference \cite{Cai1}. We further
demonstrate that CDD can be used (through the spin-locking
sequence \cite{Slichter}), to enable direct polarization of the
target nuclei \cite{Henstra1988,Reynhardt1998}.
\begin{figure}
\centering{}
\begin{minipage}[c]{85.0mm}
\begin{minipage}[c]{36mm}
\begin{center}
\includegraphics[width=35.7mm]{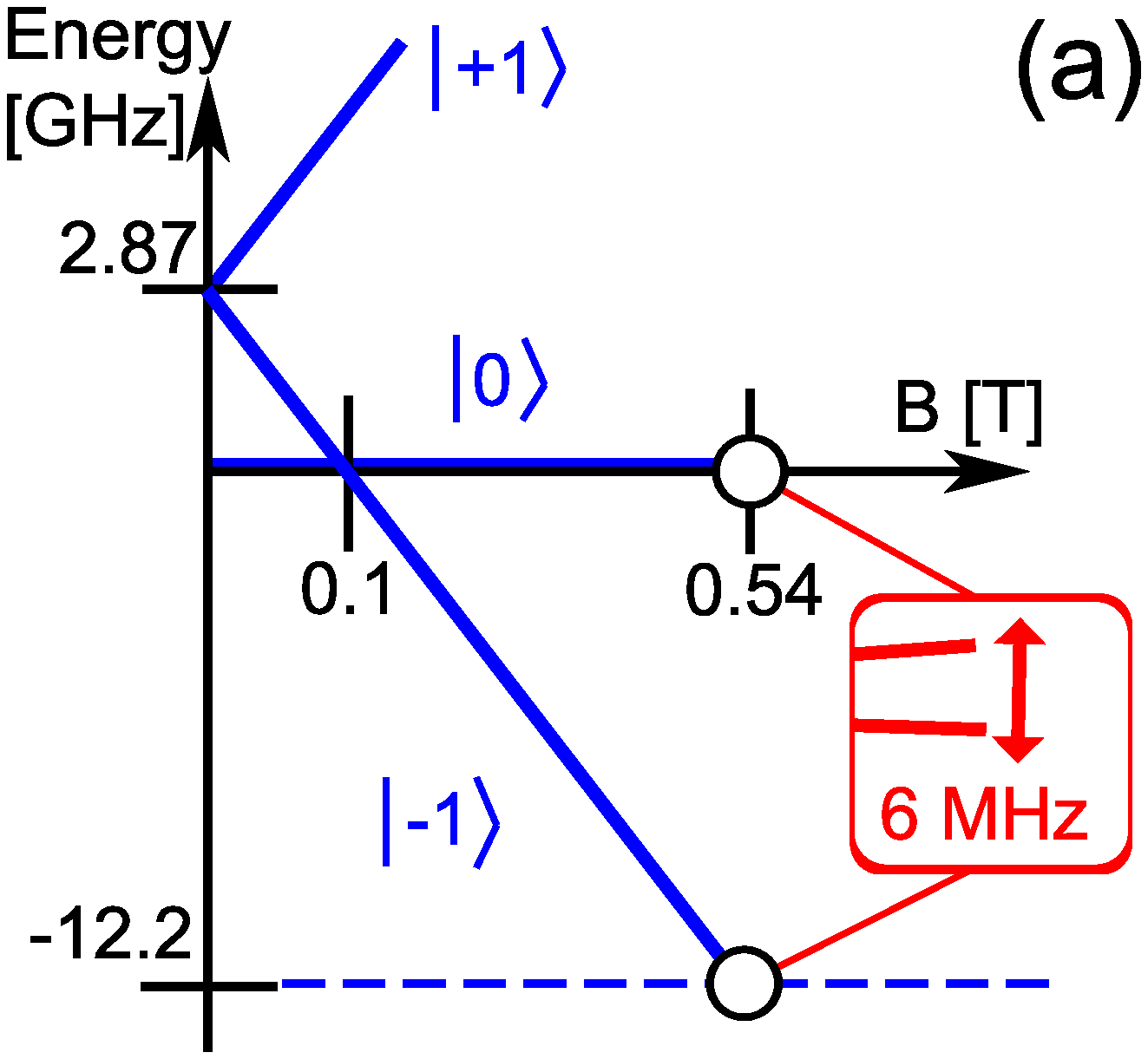}
\par\end{center}
\end{minipage}
\begin{minipage}[c]{23.2mm}
\begin{center}
\includegraphics[width=24.0mm]{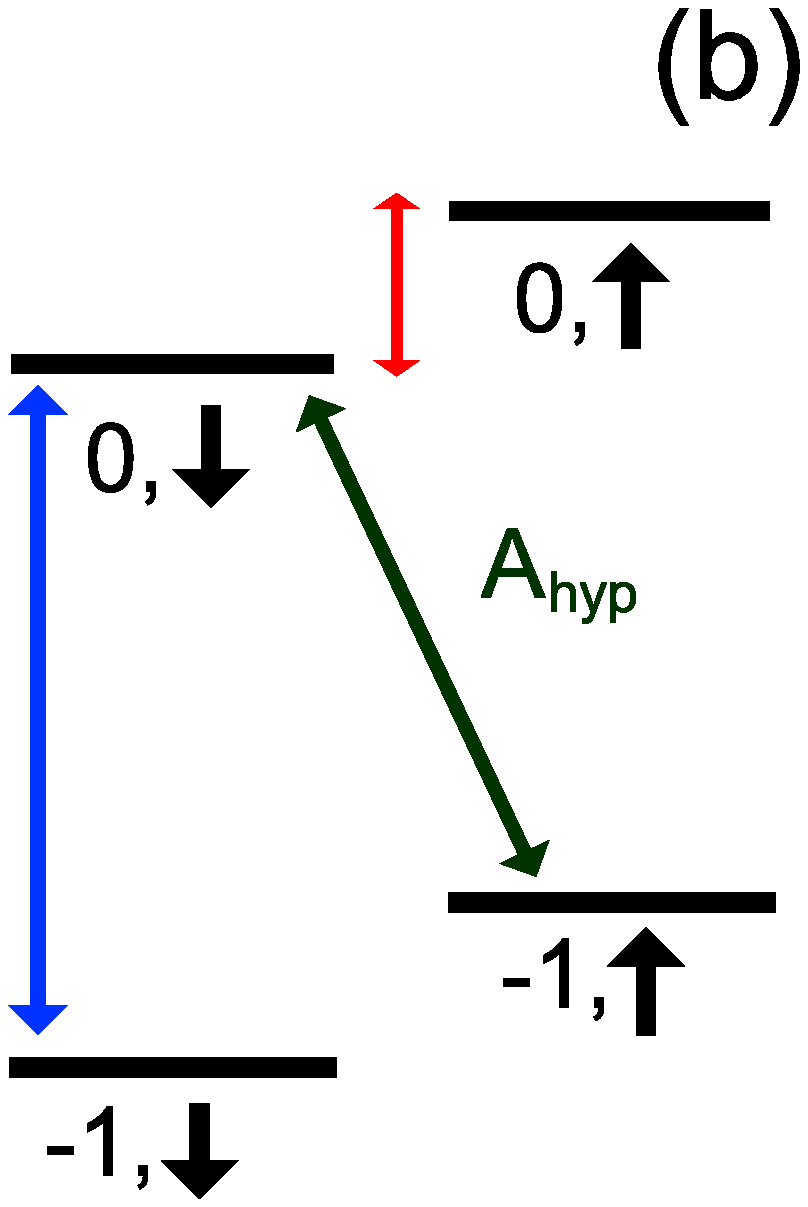}
\par\end{center}
\end{minipage}
\begin{minipage}[c]{23.2mm}
\begin{center}
\includegraphics[width=24.0mm]{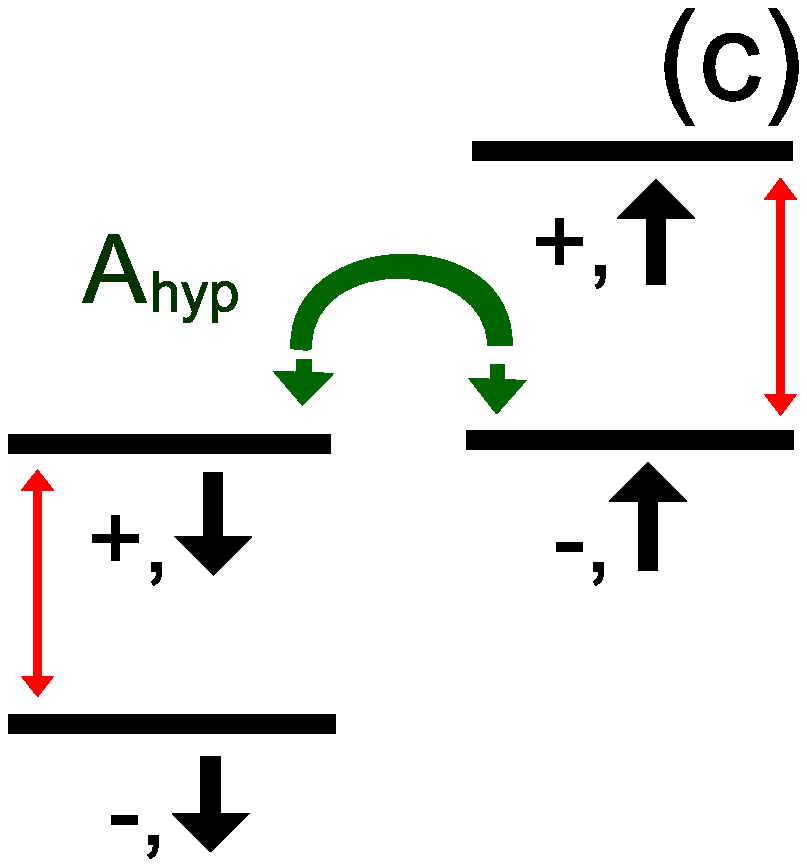}
\par\end{center}
\end{minipage}
\caption{{\small (color online) (a) Ground-state energy level
structure of the NV center as a function of an axial magnetic
field. At a magnetic field of 0.54\,T, the NV center ground states
m$_{s}$=0,-1 have an energy separation of 12 GHz, and the nuclear
system is split by 6 MHz (Red box). (b) Energy-level diagram of
the NV center electronic ground-state and a single $^{13}$C
nuclear spin. The coupling term, $\mathbf{A}_{hyp}$, induces
flip-flops between the electron spin and a coupled nuclear spin
(green arrow), which are suppressed by the mismatch of the
electronic and nuclear energies (blue and red arrows,
respectively). (c) The energy level diagram in the presence of a
resonant MW field. The electronic spin is described with the
dressed states $\left\vert \pm \right\rangle$. At HHDR, coherent
oscillations between the $\left\vert +,\downarrow \right\rangle $
level and the $\left\vert -,\uparrow \right\rangle $ level (green
curve) are enhanced .}}
\end{minipage}
\label{Figure1}
\end{figure}

HHDR occurs when two spins with distinct energy separation are
simultaneously driven so that their oscillation (Rabi) frequencies
become resonant, or alternatively, when one species is driven with
a Rabi frequency that is equal to the energy scale of the other
spin \cite{Hartmann1963}. Polarization exchange between the two
spin systems can then occur via cross-relaxation, which is usually
suppressed by their energy mismatch. In our experiments,
corresponding to the latter case, we drive a single NV electronic
spin with a Rabi frequency that matches the Zeeman energy of a
nearby nuclear spin. This enhances the coherent exchange
interaction between the two spins, which would otherwise be
prohibited due to the three orders of magnitude energy difference
(Fig. 1a,b). By adjusting the intensity of the driving field, the
NV spin sensor can be used as a tunable, narrow-band spectrometer
\cite{Cai1}, with spectral resolution limited only by the
decoupling efficiency and interrogation time.

\bigskip\emph{Hartmann-Hahn dynamics with a single NV-center}. We consider an NV electronic spin, $\mathbf{S}$,
and an additional $^{13}C$ nuclear spin, $\mathbf{I}$, with
gyromagnetic ratio $\gamma _{N}$. Their interaction can be
described by the dipole-dipole term
$H_{NV-^{13}C}=S_{z}\cdot\mathbf{A_{hyp}}\cdot\mathbf{I}$, where
$\mathbf{A_{hyp}}$ is the hyperfine vector (see \cite{Cai1,Supp}),
and non-secular terms are neglected due to the energy mismatch of
the two spins (Fig. 1a,b). In an external magnetic field
$\mathbf{B}$, the splitting between the nuclear states
$\left\vert\uparrow\right\rangle$,$\left\vert\downarrow\right\rangle$
is $\gamma_{N}\left\vert\mathbf{B_{eff}}\right\vert=\gamma
_{N}\left\vert \mathbf{B}-\left( 1/2\right)
\mathbf{A}_{hyp}\right\vert$ (Fig. 1b, red arrow). If a continuous
microwave (MW) field resonant with the $m_{s}=0,-1$ transition,
and whose intensity induces Rabi frequency, $\Omega$, is applied,
the NV center can be described by the MW-dressed states
$\left\vert \pm \right\rangle=\frac{1}{\sqrt{2}} ( \left\vert 0
\right\rangle \pm \left\vert -1 \right\rangle )$. The energy gap
of these states is $\Omega$, and an energy matching condition (the
Hartmann-Hahn condition) given by
\begin{equation}
\delta\Omega =\Omega -\gamma _{N}\left\vert
\mathbf{B_{eff}}\right\vert=0, \label{HHcondition}
\end{equation}%
can be engineered. Then, the energy of the $\left\vert
+,\downarrow \right\rangle$ state and the $\left\vert -,\uparrow
\right\rangle$ state is equal, and being coupled, they will evolve
coherently together. The remaining states $\left\vert +,\uparrow
\right\rangle$ and $\left\vert -,\downarrow \right\rangle$ are
separated by 2$\Omega$ (Fig 1c, red arrows), thus decoupled from
the joint dynamics. The probability of finding the dressed NV
center, initially set to the state $\left\vert +\right\rangle$, in
the opposite state $\left\vert -\right\rangle$, after time $\tau$,
is
\begin{equation}
p\left( \tau,\delta\Omega \right)=\frac{J^{2}}{J^{2}+\delta\Omega^{2}} \times \sin^{2} \left( \sqrt{ %
J^{2} + \delta\Omega^{2} } \frac{\tau}{2} \right),
\label{EqProbabilty}
\end{equation}%
where $J$, given by
\begin{equation}
J=\frac{1}{4}\gamma _{N}\left\vert \mathbf{A}_{hyp}\right\vert
\sin \theta, \label{eqJcoup}
\end{equation}%
is proportional to the coupling strength, and depends on $\theta$,
 the angle between $\mathbf{B_{eff}}$ and $\mathbf{A}_{hyp}$
(see \cite{Supp}). The transition probability (Eq.
(\ref{EqProbabilty})) shows temporal oscillatory behavior, and a
spectral dependence (Lorentzian shape of width $J$). The former is
a manifestation of the coherent nature of this interaction:
starting in the $\left\vert +,\downarrow \right\rangle $ state,
the system evolves according to $\left\vert \Psi \right\rangle
=\left\vert +,\downarrow \right\rangle \cos \left( Jt\right)
+\left\vert -,\uparrow \right\rangle \sin \left( Jt\right)$. Thus,
at time $t=\pi /2J$ the two spins become maximally entangled, and
after a $t=\pi /J$ a full population transfer occurs; i.e. the
states of the two spins are swapped. The latter spectral
dependence in Eq. (\ref{EqProbabilty}) reflects that coherent
oscillations between the NV center and weakly coupled nuclear
spins are extremely sensitive to detuning from the Hartmann-Hahn
condition.

\bigskip\emph{Single nuclear spin spectroscopy and imaging}. In our
experiments, HHDR is performed with single NV centers in a natural
abundance ($ ^{13}$C 1.11\%) diamond. (Details on the diamond
sample, and on the experimental setup and methods can be found in
the supplementary material \cite{Supp}). In order to increase the
decoupling efficiency, we apply a high Rabi frequency of
$\sim$\,6\,MHz which is matched by the Larmor frequency of the
$^{13}$C nuclear spins in a magnetic field of 0.54\,T (Fig. 1a).
The transition probability in Eq.(\ref{EqProbabilty}) can be
measured in a straightforward way by applying a spin-locking
sequence \cite{Hartmann1963,Henstra1988}. In this sequence, the NV
electronic spin is first optically polarized by 532nm light
illumination. Then rotated to the $\left\vert +\right\rangle$
state with a $\pi$/2 pulse, and maintained there with a continuous
driving field applied at the same frequency, but with a
90$^{\circ}$ phase shift \cite{Supp}. Away from HHDR, the NV spin
remains in the $\left\vert +\right\rangle$ state and is
subsequently rotated to the low fluorescent, $\left\vert
-1\right\rangle$ state by a final pi/2 pulse (Fig. 2a). However at
the resonance condition, the NV spin undergoes coherent
oscillations between the $\left\vert +\right\rangle$ and
$\left\vert -\right\rangle$ states. After the second $\pi$/2 pulse
this evolution is observed as modulations in the NV fluorescence
as the final state oscillates between the $\left\vert
0\right\rangle$ and $\left\vert -1\right\rangle$ states. As
described later, this protocol produces polarization of the
nuclear bath, inhibiting further interaction between the NV center
and nearby nuclear spins. Therefore, for spectroscopy
measurements, i.e characterization of the coupling strength and
orientation, an alternating version of the spin-locking sequence
was used (Fig. 2b), which produces the same experimental signal
but without polarization of the nuclear bath. It comprises two
similar sequences \textquotedblleft+\textquotedblright and
\textquotedblleft-\textquotedblright, essentially initializing the
NV into the $\left\vert +\right\rangle$ and $\left\vert
-\right\rangle$ states, respectively. These induce nuclear
polarization in alternating directions, thus the net nuclear
polarization is zero.

\begin{figure}
\begin{minipage}[c]{85.0mm}
\includegraphics[width=76mm,height=32.4mm]{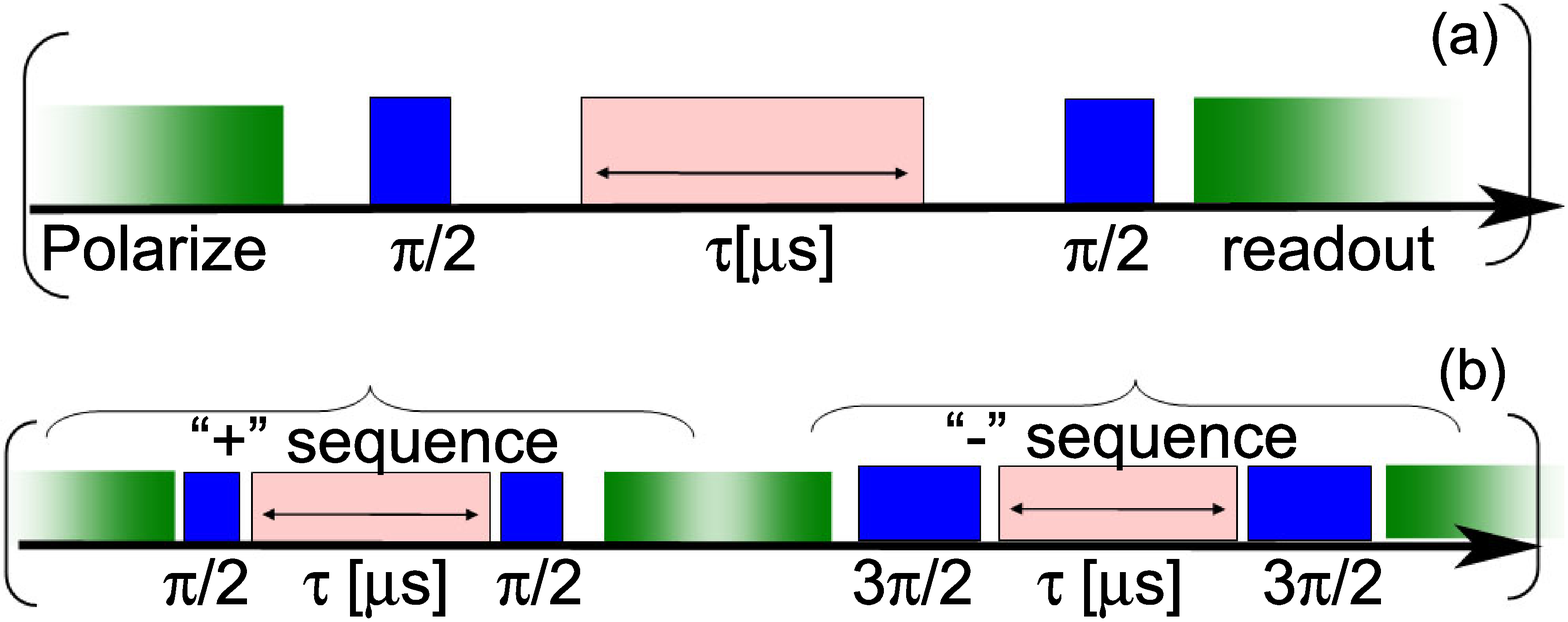}
\includegraphics[width=76mm,height=30mm]{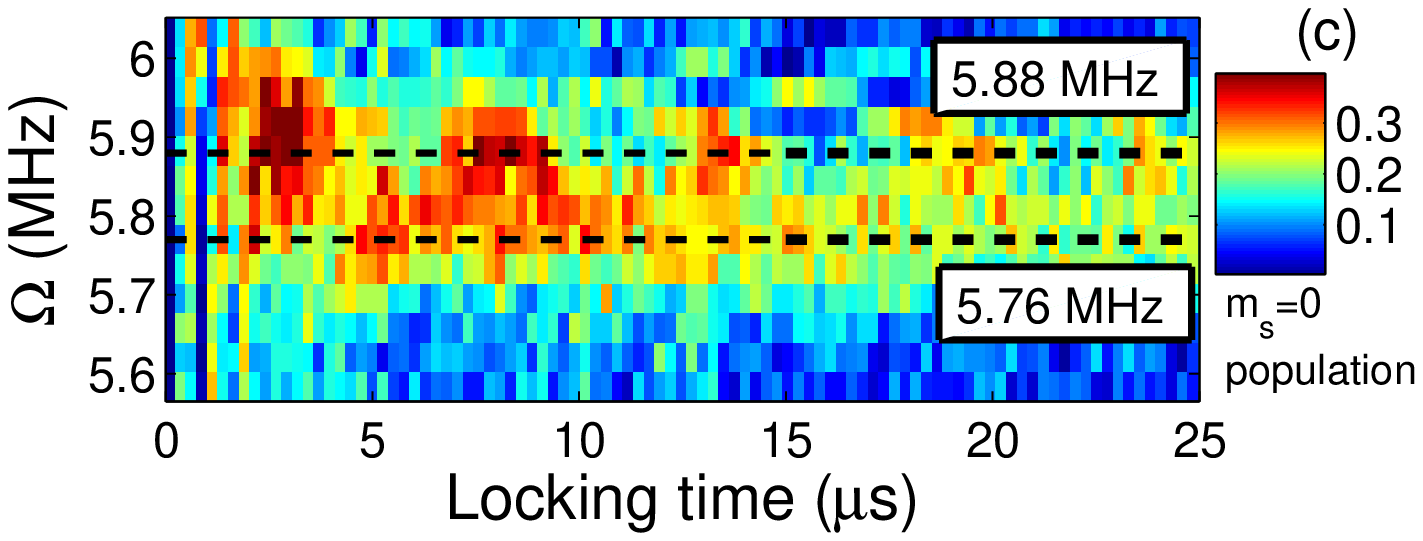}
\includegraphics[width=76mm,height=35mm]{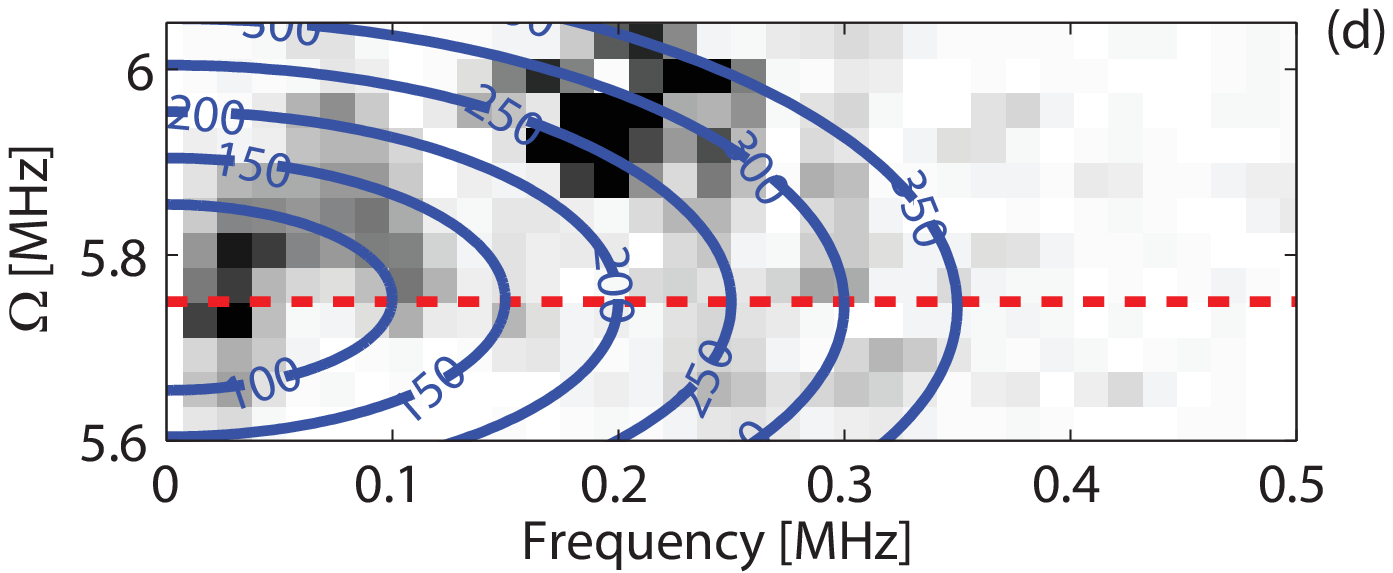}
\caption{{\small (Color online) Weakly coupled nuclear spin
spectroscopy. (a) The spin-locking sequence; 532nm-light pulses
are marked in green, MW pulses in blue and pink, for X and Y
pulses, respectively. (b) alternating spin-locking sequence. (c)
Experimentally observed population of the $m_s=$0 state as a
function of the MW driving field $\Omega$ and the spin-locking
time $\tau$. Colormap is normalized by the contrast observed in a
Rabi experiment (performed with a strong driving, $\Omega>13$MHz).
The black dashed lines are guide to the eye. (d) Fourier analysis
of the spin-locking signal for various MW driving fields $\Omega$.
Coupling to a single nuclear spin is apparent as a dark spot
around 200 kHz, while coupling to the bath produces the lower
frequency signal. The blue lines correspond to the predicted
values of the resonance driving amplitude $\gamma
_{^{13}C}\left\vert\mathbf{B_{eff}}\right\vert$
(Eq.(\ref{HHcondition})) and the flip-flop rate $J$
(Eq.(\ref{eqJcoup})), respectively, for a constant coupling
strength and various angles (values are in kHz). The red line is
the value of the \textquotedblleft bare\textquotedblright
 Hartmann-Hahn term $\Omega=\gamma _{^{13}C}\left\vert\mathbf{B}\right\vert $.}}
\end{minipage}
\end{figure}

Fig. 2c shows the transition probability of a single NV center
interacting with the surrounding spin-bath. Two features which
correspond to interaction with nuclear spins can be seen at
$\Omega \simeq5.76$\,MHz and at $\Omega \simeq5.9$\,MHz. The first
agrees well with the expected Larmor frequency for $^{13}$C spins
in the applied field ($\left\vert \mathbf{B}\right\vert=5375$\,G),
and shows loss of coherence of the $\left\vert +\right\rangle$
state due to the interaction with many nuclear spins. The second
feature is the realization of HHDR with a single nuclear spin,
whose coupling strength with the NV center ($\sim$200 kHz) is 2.5
times smaller than the measured inhomogeneous ($1/T_2^*$)
linewidth, which characterizes the phase-detection sensitivity
without decoupling. Note, Eq.\,(\ref{EqProbabilty}) neglects the
NV electronic spin interaction with its host nitrogen nuclear spin
($^{15}$N in this case). For our experimental parameters, HHDR is
efficient for a single hyperfine projection which has a time
averaged population of 0.45 \cite{Supp}. Therefore, the 40\%
oscillation contrast indicates $\sim$90\% polarization exchange
efficiency with the single $^{13}$C spin. The two-dimensional
nature of Eq.\,(\ref{EqProbabilty}), i.e. the spectral and
temporal dependencies, also allows for nuclear spin imaging (Fig.
2d). Both the optimal Rabi frequency $\Omega_{opt}$ which
satisfies Eq.\,(\ref{HHcondition}), and the oscillation rate at
double-resonance (Eq.\,(\ref{eqJcoup})) contain information about
the interaction strength, $\left\vert \mathbf{A}_{hyp}\right\vert$
and its orientation, $\theta$. Inverting Eq.\,(\ref{HHcondition})
and Eq.\,(\ref{eqJcoup}) for this electron-nuclear pair
($\Omega_{opt}=2\pi\times\left(5.88\pm0.03\right)$\,MHz,
($J=2\pi\times\left(188\pm30\right)$\,kHz), we deduce that the
coupling for this pair is $(1/4)\gamma _{N}\left\vert
\mathbf{A}_{hyp}\right\vert = 2\pi \times 220\pm $40\,kHz (which
corresponds to a nuclear spin located $\sim$\,0.5\,nm from the NV
center, assuming the contribution from the contact term in the
interaction is negligible), and the orientation is
$\theta=56\pm10^{\circ}$.

\begin{figure*}
\centering{}
\begin{minipage}[t]{18.0cm}
\begin{minipage}[c]{59.0mm}
\begin{center}
\includegraphics[width=59mm,height=44.25mm]{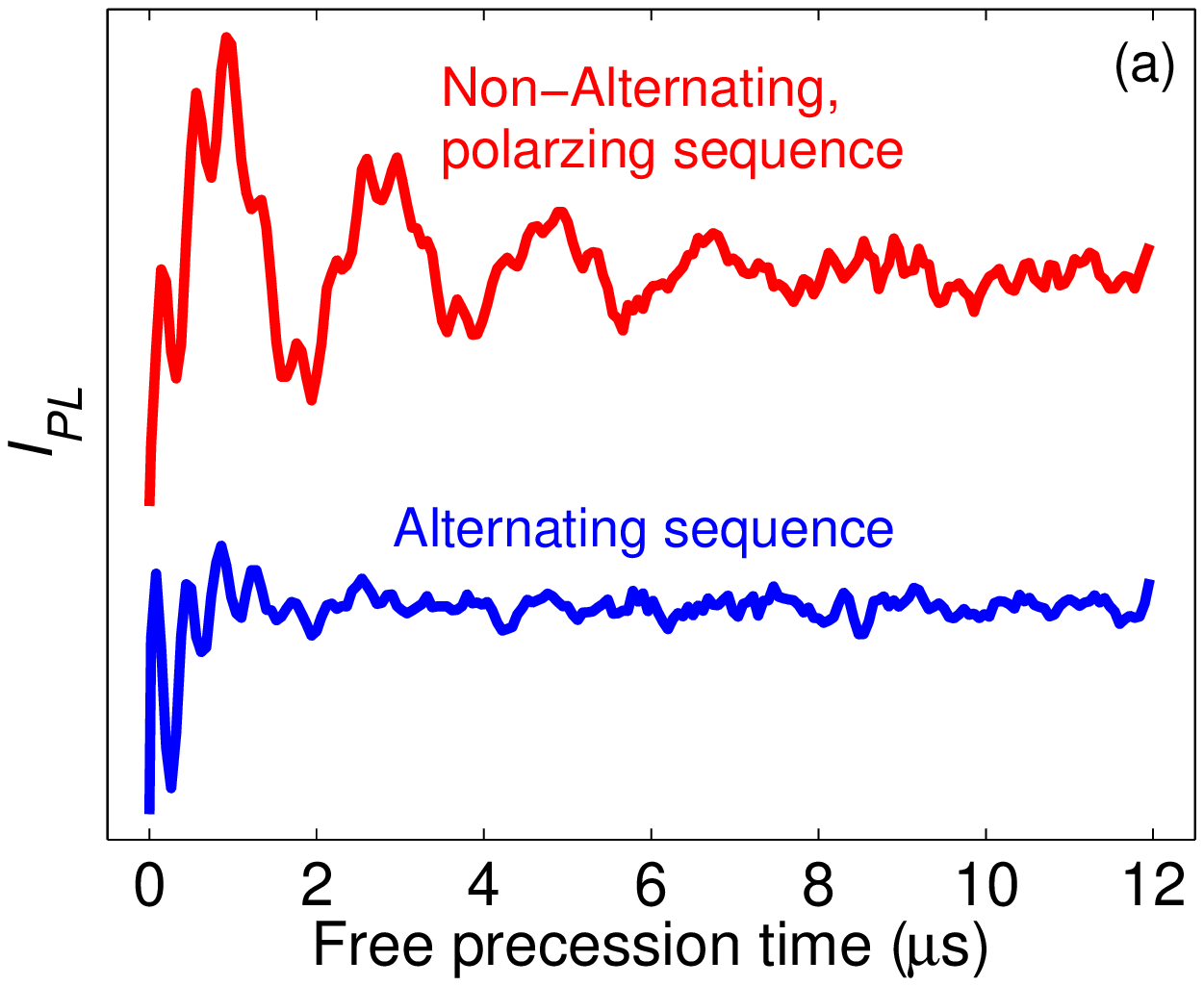}
\par\end{center}
\end{minipage}
\begin{minipage}[c]{59.0mm}
\begin{center}
\includegraphics[width=59.0mm,height=44.25mm]{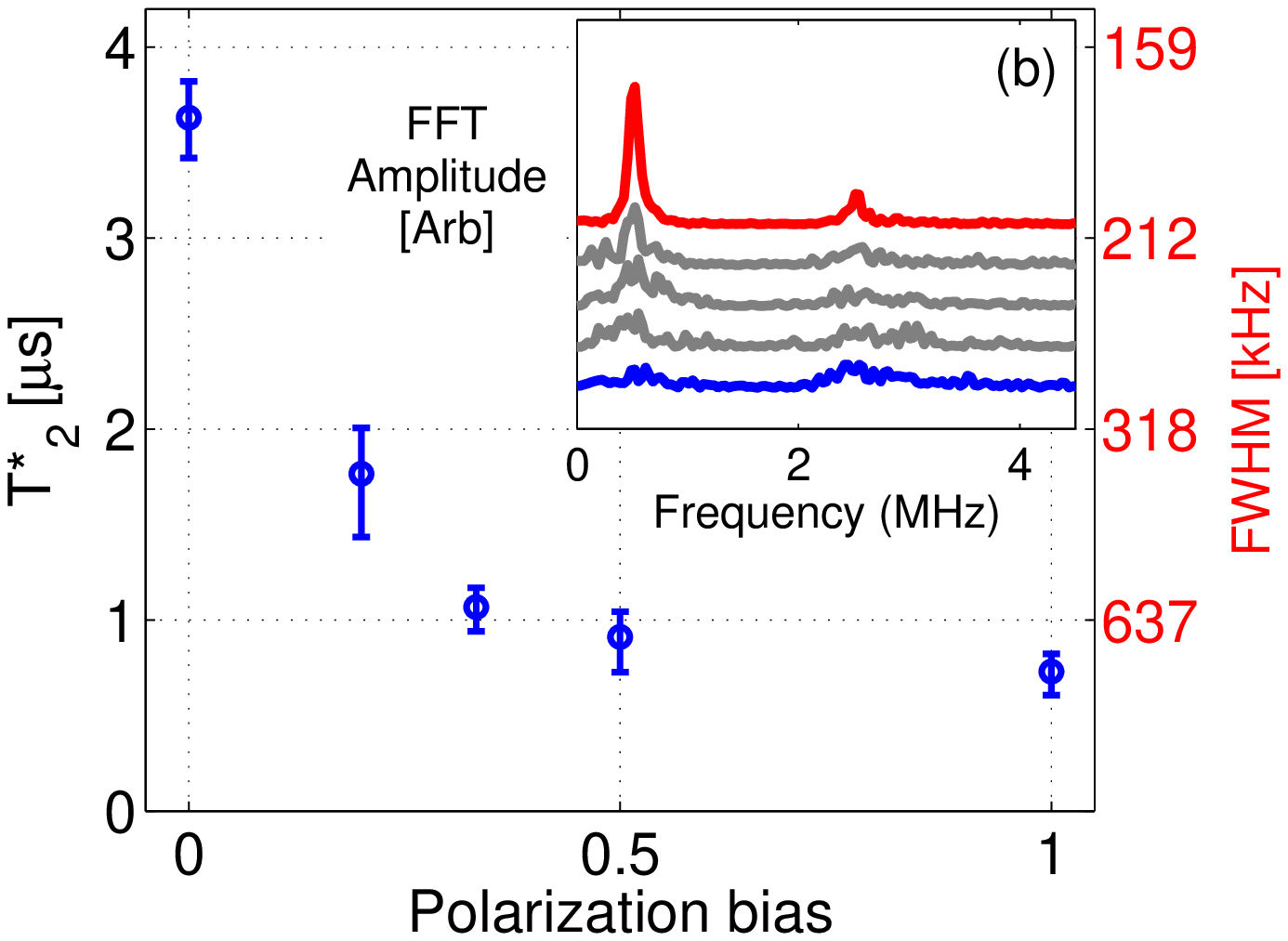}
\par\end{center}
\end{minipage}
\begin{minipage}[c]{59.0mm}
\begin{center}
\includegraphics[width=59.0mm,height=44.25mm]{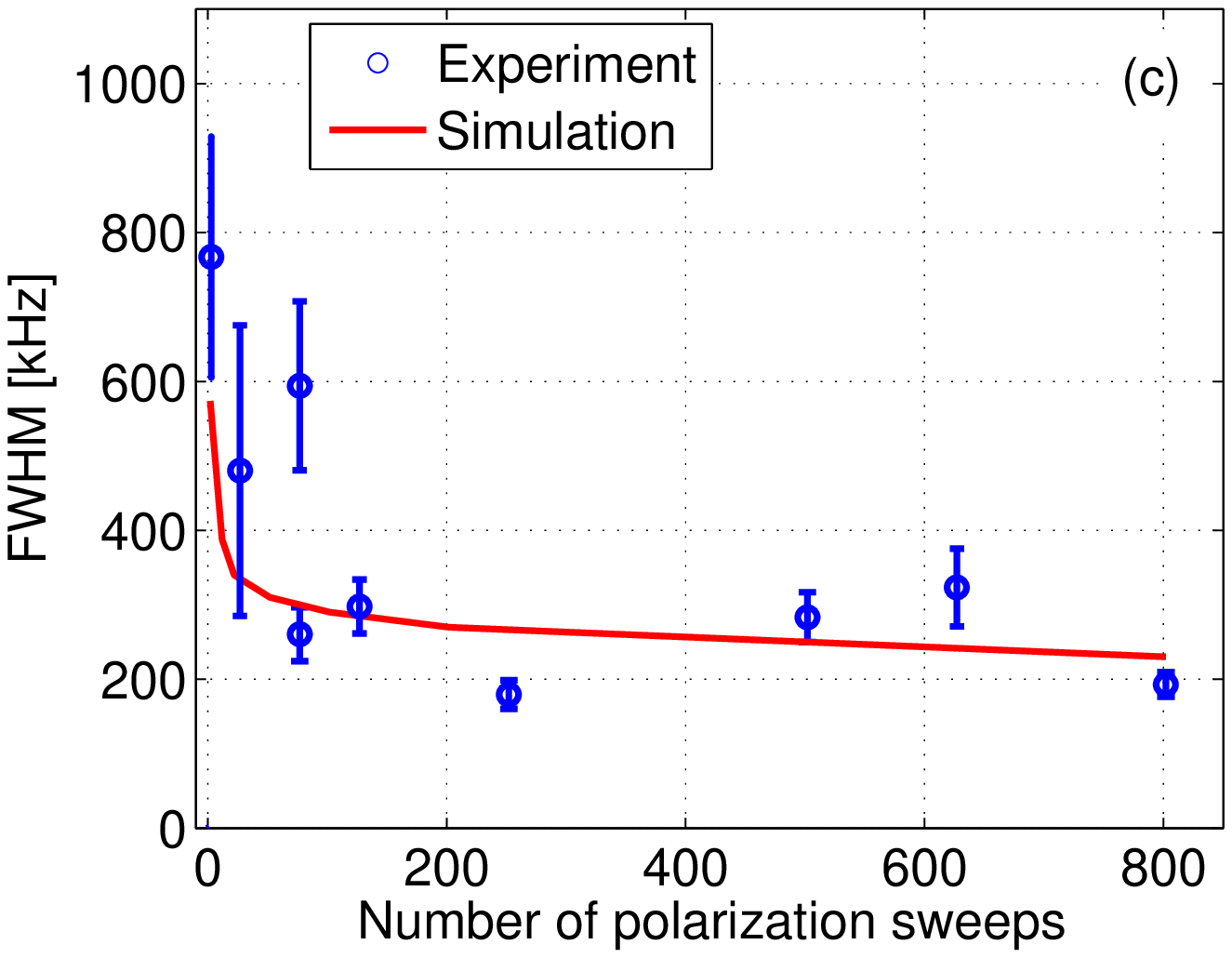}
\par\end{center}
\end{minipage}
\caption{(color online) Dynamical polarization of the nuclear
spin-bath. (a) Free induction decay measured on a single NV center
while applying the alternating sequence (blue curve -
$T_{2}^{\star}\sim$0.6 $\,\mu$s) and while applying the
polarization sequence (red curve - $T_{2}^{\star}\sim$3 $\,\mu$s).
(b) Dependence of the dephasing time $T_{2}^{\star}$ on the
polarization bias. The inset shows the Fourier amplitudes of the
corresponding FID signals, showing a monotonic narrowing as the
bias is increased. The lower blue curve and the upper red curves
correspond to polarization balance of 1 and 0, respectively, as
plotted in Fig. 3a. (c) Build up of the nuclear spin-bath
polarization is reflected in a narrowing of the full-width at
half-maximum (FWHM) linewidth in the Ramsey measurements as the
number of polarization sweeps is increased. The experimental data
(circles) qualitatively agree with simulation of direct
polarization mechanism under the spin-temperature approximation
(red solid line). The simulation includes an additional offset of
150\,kHz to account for magnet drifts on the linewidth. In (b),(c)
the $T_{2}^{\star}$ and FWHM values are extracted by fitting the
function Y$ = \Lambda_{1} $ exp$(-((f-\mu 1)/ \sigma)^{2})+
\Lambda_{2}$ exp$(-((f- \mu_{2})/\sigma)^2)$ to the Fourier
transform spectra (Fig. 3b inset), and by
FWHM$=2\sigma\sqrt{\ln{2}}$ and $T_{2}^{\star}=2/$($\pi$ FWHM).
The error bars are one standard deviation obtained from the fit
process.}
\end{minipage}
\label{Figure3}
\end{figure*}

The measured coupling, $\sim$\,200\,kHz, does not mark the
ultimate sensitivity of our scheme. Coherent oscillations of the
NV-nuclear pair last for more than 25\,$\mu$s, implying that a
40\,kHz coupling could have been detected if such a nuclear spin
was present in the vicinity of this NV center, and providing it
could be spectrally separated from the spin bath signal at
$5.76$\,MHz ($\gamma _{^{13}C}\left\vert\mathbf{B}\right\vert $).
In principle, the interrogation time and hence the sensitivity of
the HHDR scheme, is limited by $T_{1}^{\rho}$ - the longitudinal
relaxation time of the NV center in the rotating frame
\cite{Slichter}. $T_{1}^{\rho}$ times exceeding one millisecond
have been measured for NV centers at room temperature
\cite{Naydenov2011}, which translates to sub-kHz resolution.
However both practical and fundamental aspects limit the
sensitivity of the scheme. First, fluctuations in the applied MW
and static magnetic fields cause broadening and reduce the
achieved interrogation time. This may be overcome with improved
concatenated continuous driving schemes which mitigate the impact
of MW instabilities \cite{Cai2}. Second, the decoupling efficiency
of CDD depends on the spectral overlap of the environmental noise
spectrum with the decoupling filter function \cite{Taylor2008}.
The overlap may be reduced by modifying either the filter function
or the bath spectrum to achieve optimal decoupling performance.
For example, to target detection of protons
\cite{Mamin2013,Staudacher2013}, the NV-center can be tuned to the
proton spectral region, which is detuned from $^{13}$C nuclear
spins in moderate magnetic fields. However, in our experiment we
aimed to separate the signal of individual $^{13}$C nuclear spins
from a bath comprised from the same nuclear species. Then the
spectral density of the bath is peaked near the interrogated
frequencies (shifted by only the coupling interaction between the
sensor and target spin) leading to a reduced $T_{1}^{\rho}$
coherence time \cite{vanOort1989}. For a detailed discussion on
the sensitivity of the scheme, see
\cite{Degen2013,Cappellaro2012}, and \cite{Supp}.

\bigskip\emph{Nuclear spin-bath polarization}. In addition to the
detection of single or few nuclear spins, one can utilize the
direct flip-flops between the NV center and nuclear spins to
polarize the surrounding bath (Figure 3). Under HHDR, the
$\left\vert +,\downarrow \right\rangle \longleftrightarrow
\left\vert -,\uparrow \right\rangle $ transition allows transfer
of polarization from the NV electronic spin to resonant nuclear
spins. Therefore, when optical polarization of the NV spin is
established at the beginning of each sweep an efficient cooling
mechanism of the nuclear spin-bath is provided. We note that other
transitions between the dressed-electronic spin and the nuclear
spin can lead to a reversal of polarization \cite{Supp}. However
these transitions are suppressed by an energy mismatch described
by $\sim(J/\Omega )^{2}$, and in our high-field experiments are of
the order $1\times 10^{-3}$. We observe the bath polarization
experimentally in the spin-locking signal when employing the
non-alternating sequence (Fig. 2a), which shows no oscillations as
the system is driven into a non-interacting state in which all the
nuclear spins are polarized to their up state \cite{Supp}. The
bath polarization itself can also be directly observed from the
free induction decay (FID) signal of the NV center, measured using
a Ramsey sequence.

The results show that when the bath is polarized the NV phase
memory time, $T_{2}^{\star}$, increases five-fold in comparison
with a non-polarized bath (Fig. 3a). Further improvements in
$T_{2}^{\star}$ are limited by magnet drifts of our setup. To
investigate the polarization dynamics further, the polarization
rates towards the up and down state were balanced using N$_{+}$
sweeps of the \textquotedblleft+\textquotedblright sequence and
N$_{-}$ sweeps of the \textquotedblleft-\textquotedblright
sequence. We define the polarization \emph{bias} as
(N$_{+}$-N$_{-}$)/(N$_{+}$+N$_{-}$). The smooth transition of
$T_{2}^{\star}$ times in the range 0.6--3\,$\mu$s when adjusting
the polarization bias from zero to unity indicates that precise
control over spin-bath degree of polarization is achievable (Fig.
3b). Finally, we measure the dynamics of the bath polarization by
varying the number of polarization sweeps and measure the FID
signal (Fig. 3c) \cite{Comment2}. The experimental results are in
qualitative agreement with a numerical simulation of a master
equation for a single NV center surrounded by 500 $^{13}$C spins
\cite{Christ2007}, showing a characteristic gradual polarization.
The simulation indicates that close lying nuclear spins are
polarized very efficiently (almost one spin per sweep), whereas
farther away nuclear spins are polarized much slower \cite{Supp}.
The proximal spins have the greatest influence on the FID
linewidth, thus their polarization improves $T_{2}^{\star}$
significantly. However this also creates an inherent problem when
comparing numerical simulations to the experiment, as both are
dependent on the actual configuration of nearby nuclei.
Initializing and probing the nuclear-bath state as demonstrated
here provides a route for characterizing fundamental processes
such as inter-nuclear interactions. For example, it is of great
interest to discriminate the aforementioned direct polarization
process from spin-diffusion induced polarization process
\cite{Fischer2012}.

\bigskip
\emph{Conclusions}. Continuous dynamical decoupling allows a
single NV-center to sense minute magnetic fields originating from
a single nuclear spin, in spite of the large background noise
produced by its environment. We demonstrated that a careful tuning
of the protocol may bring forth room-temperature hyperpolarization
among nuclei in the surrounding bath. The interaction between the
NV electronic spin and the nuclear spins preserves its coherent
nature, i.e. it can support quantum information protocols using
dressed qubits \cite{BermudezPRL2011}.

In biological measurements which are characterized with an
extremely disruptive environment, CDD can become an optimal tool:
First, it allows improved decoupling through high Rabi frequencies
at efficient energy expenditure compare to pulsed techniques, and
would thus be less invasive to biological samples. Second, using
room-temperature nuclear polarization the target spin signal can
be amplified, resulting with signal-to-noise ratio improvement
according to $\sqrt{N}$ where $N$ is the number of nuclear spins.
Moreover, for many diamond-based QIP protocols, initialization of
the nuclear bath to a given state is essential, for example in
quantum simulators \cite{Cai3}. We also note that Hartmann-Hahn
double resonance can be applied for the detection of electron
spins as was demonstrated recently \cite{Wals}.

\bigskip

The authors thank Rainer Pfeiffer and Kay Jahnke for assistance
with the experiments. The authors are grateful to Philip Hemmer,
J\"{o}rg Wrachtrup, Vyacheslav Dobrovitskii and Philipp Neumann
for fruitful discussions. The research was supported by DFG
(FOR1482, SPP1601  and SFB TR21), EU (DIAMANT), DARPA (QUASAR) and
the Alexander von Humboldt Foundation. J.-M.C acknowledges the
support of Marie-Curie fellowship and FP7.


\begin{thebibliography}{99}
\bibitem[29]{Comment2} To initialize (depolarize) the nuclear bath before each measurement, we have used many sweeps of the alternating sequence.
\end{thebibliography}
\end{document}